# Spectroscopic Needs for Calibration of LSST Photometric Redshifts

**A white paper submitted to the National Research Council**

**Committee on a Strategy to Optimize the U.S. O/IR System in the Era of LSST**

Samuel J. Schmidt[1], Jeffrey A. Newman[2], Alexandra Abate[3], Filipe B. Abdalla[4], Sahar Allam[5], Steven W. Allen[6,7], Réza Ansari[8], Stephen Bailey[9], Wayne A. Barkhouse[10], Timothy C. Beers[11], Michael R. Blanton[12], Mark Brodwin[13], Joel R. Brownstein[14], Robert J. Brunner[15], Matias Carrasco Kind[15], Jorge L. Cervantes-Cota[16], Elliott Cheu[3], Nora Elisa Chisari[17], Matthew Colless[18], Johan Comparat[19], Jean Coupon[20], Carlos E. Cunha[21], Axel de la Macorra[22], Ian P. Dell'Antonio[23], Brenda L. Frye[24], Eric J. Gawiser[25], Neil Gehrels[26], Kevin Grady[26], Alex

---

[1] Dept. of Physics, University of California, One Shields Ave., Davis, CA 95616, USA
[2] Department of Physics and Astronomy and PITT PACC, University of Pittsburgh, 3941 O'Hara St., Pittsburgh, PA 15260
[3] Physics Department, University of Arizona, 1118 East 4th Street, Tucson, AZ 85721, USA
[4] Astrophysics Group, Department of Physics & Astronomy, University College London, Gower Place, London WC1E 6BT, UK
[5] Fermi National Accelerator Laboratory, MS 127, PO Box 500, Batavia, Illinois 60510, USA
[6] Dept. of Physics, Stanford University, 382 Via Pueblo Mall, Stanford, CA 94305, USA
[7] SLAC National Accelerator Laboratory, 2575 Sand Hill Road MS 29, Menlo Park, CA 94025, USA
[8] Université Paris-Sud, LAL-IN2P3/CNRS, BP 34 , 91898 Orsay Cedex, France
[9] Lawrence Berkeley National Laboratory, 1 Cyclotron Rd, Berkeley, CA 94720, USA
[10] Department of Physics and Astrophysics, University of North Dakota, Grand Forks, ND 58202, USA
[11] National Optical Astronomy Observatories, 50 N. Cherry Avenue, P.O. Box 26732, Tucson, AZ 85726, USA
[12] Department of Physics, New York University, 4 Washington Place, Room 424, New York, NY 10003, USA
[13] Department of Physics and Astronomy, University of Missouri at Kansas City, 5110 Rockhill Road, Kansas City, MO 64110, USA
[14] Department of Physics, The University of Utah, 115 S 1400 E, Salt Lake City, UT 84112, USA
[15] Dept. of Astronomy, University of Illinois, 226 Astronomy Building, MC-221, 1002 W. Green St., Urbana, IL 61801, USA
[16] Instituto Nacional de Investigaciones Nucleares (ININ), Apartado Postal 18-1027 Col. Escandón, México DF 11801, México
[17] Department of Astrophysical Sciences, Princeton University, 4 Ivy Lane, Peyton Hall, Princeton, NJ 08544, USA
[18] Research School of Astronomy and Astrophysics, The Australian National University, Canberra, ACT 2611, Australia
[19] Campus of International Excellence UAM+CSIC, Cantoblanco, E-28049 Madrid, Spain
[20] Astronomical Observatory of the University of Geneva, ch. d'Ecogia 16, 1290 Versoix, Switzerland
[21] Kavli Institute for Particle Astrophysics and Cosmology, 452 Lomita Mall, Stanford University, Stanford, CA 94305, USA
[22] Depto. de Fisica Teorica and Instituto Avanzado de Cosmologia (IAC), UNAM, Mexico City, Mexico
[23] Department of Physics, Brown University, Box 1843, 182 Hope Street, Providence, RI 02912, USA
[24] Department of Astronomy and Steward Observatory, University of Arizona, 933 North Cherry Avenue, Tucson, AZ 85721, USA
[25] Dept. of Physics & Astronomy, Rutgers, The State University of New Jersey, 136 Frelinghuysen Rd., Piscataway, NJ 08854, USA
[26] NASA Goddard Space Flight Center, Greenbelt, MD 2077, USA

Hagen[27], P. Hall[28], Andew P. Hearin[29], Hendrik Hildebrandt[30], Christopher M. Hirata[31], Shirley Ho[32], Klaus Honscheid[33], Dragan Huterer[34], Željko Ivezić[35], Jean-Paul Kneib[36,37], Jeffrey W. Kruk[26], Ofer Lahav[4,] Rachel Mandelbaum[32], Jennifer L. Marshall[38], Daniel J. Matthews[2], B. Ménard[39], Ramon Miquel[40], Marc Moniez[8], H. W. Moos[39], John Moustakas[41], Adam D. Myers[42], C. Papovich[38], John A. Peacock[43], Changbom Park[44], Mubdi Rahman[39], Jason Rhodes[45], Jean-stephane Ricol[46], Iftach Sadeh[4], Anže Slozar[47], Daniel K. Stern[45], J. Anthony Tyson[1], A. von der Linden[6], R. Wechsler[6,7], W. M. Wood-Vasey[2], Andrew R. Zentner[2]

E-mail Addresses: sschmidt@physics.ucdavis.edu; janewman@pitt.edu (corresponding author); abate@email.arizona.edu

**Context:** This white paper is a summary of the photo-z calibration needs described in the Snowmass White Paper *Spectroscopic Needs for Imaging Dark Energy Experiments* by Newman et al., available at http://arxiv.org/abs/1309.5384. That white paper focuses on estimating the amount of spectroscopic redshift data required to enable photometric redshift measurements with future imaging dark energy surveys. It divides the applications of spectroscopy into *training*, i.e., the use of spectroscopic redshifts to improve algorithms and reduce photo-z errors; and *calibration*, i.e., the accurate characterization of biases and uncertainties in photo-z's, which is critical for dark energy inference. We summarize here the conclusions from that white paper relevant to the calibration of LSST photometric redshifts; a separate white paper (Abate et al.) will focus on training needs. We refer the reader to the Snowmass white paper for all references. The problem of *calibration* is that of determining the true overall redshift distribution of samples of objects selected in some way; miscalibration will lead to systematic errors in photo-z's and hence in almost all extragalactic science with LSST. For LSST, for instance, it is estimated that the mean redshift for each sample used for cosmology (typically, objects selected within some bin in photometric redshift) must be known to $\sim 2\times 10^{-3}(1+z)$, i.e., 0.2%.

---

[27] Department of Astronomy & Astrophysics, The Pennsylvania State University, 525 Davey Lab, University Park, PA 16802, USA
[28] Department of Physics and Astronomy, York University, 4700 Keele Street, Toronto, ON, Canada
[29] Yale Center for Astronomy & Astrophysics, Yale University, New Haven, CT
[30] Argelander-Institut für Astronomie, Auf dem Hügel 71, 53121 Bonn, Germany
[31] Department of Astronomy, The Ohio State University, 140 West 18th Avenue, Columbus, OH 43210, USA
[32] McWilliams Center for Cosmology, Carnegie Mellon University, 5000 Forbes Avenue, Pittsburgh, PA 15213, USA
[33] Department of Physics, The Ohio State University, 140 West 18th Avenue, Columbus, OH 43210, USA
[34] Department of Physics, University of Michigan, 450 Church St., Ann Arbor, MI 48109, USA
[35] Astronomy Department, University of Washington, PAB 357, 3910 15th Ave NE, Seattle, WA, USA
[36] Laboratoire d'Astrophysique, Ecole Polytechnique Fédérale de Lausanne (EPFL), Observatoire de Sauverny, CH-1290 Versoix, Switzerland
[37] Laboratoire d'Astrophysique de Marseille - LAM, Université d'Aix-Marseille & CNRS, UMR7326, 38 rue F. Joliot-Curie, 13388 Marseille Cedex, France
[38] Department of Physics & Astronomy, Texas A&M University, College Station, TX 77843, USA
[39] Department of Physics and Astronomy, Johns Hopkins University, Baltimore, MD 21218, USA
[40] Institut de Fisica d'Altes Energies (IFAE), Edifici Cn, Universitat Autonoma de Barcelona, E-08193 Bellaterra (Barcelona) Spain
[41] Department of Physics and Astronomy, Siena College, 515 Loudon Road, Loudonville, NY 12211, USA
[42] Department of Physics & Astronomy, University of Wyoming, 1000 E. University, Dept. 3905, Laramie, WY 82071, USA
[43] Institute for Astronomy, University of Edinburgh, Royal Observatory, Edinburgh EH9 3HJ, UK
[44] School of Physics, Korea Institute for Advanced Study, 85 Hoegiro, Dongdaemun-gu, Seoul 130-722, Korea
[45] Jet Propulsion Laboratory, California Institute of Technology, 4800 Oak Grove Drive, Pasadena, CA 91109, USA
[46] Laboratoire de Physique Subatomique et de Cosmologie Grenoble, 53 rue des Martyrs, 38026 Grenoble Cedex, France
[47] Brookhaven National Laboratory, P.O. Box 5000, Upton, NY 11973-5000

If extremely high completeness (>99.9%) is attained in the training samples described in the Abate et al. white paper, then LSST calibration requirements would be met with no need for additional work. However, since existing deep redshift samples have failed to yield secure redshifts for a systematic 20%-60% of their targets, it is a strong possibility that future deep O/IR redshift samples will not solve the calibration problem. The best options in this scenario are provided by cross-correlation methods. These techniques take advantage of the fact that bright galaxies (whose spectroscopic redshifts may be measured easily) and fainter objects (which can only be studied with photo-z's) both trace the same underlying dark matter distribution. These methods cross-correlate positions on the sky of objects with known redshifts with the locations of those galaxies whose redshift distribution we aim to characterize. By measuring this signal as a function of the known spectroscopic redshift, one can determine the *z* distribution of a purely photometric sample with high accuracy.

**Basic Requirements:** Cross-correlation calibration for LSST will require spectroscopy of a minimum of $\sim 10^5$ objects (in order to limit shot noise) spanning hundreds of square degrees (which limits the impact of field-to-field variations in measured clustering amplitudes, which dominates errors if field sizes are small). The autocorrelation properties of the spectroscopic sample must also be measured, requiring that, if spectroscopic data is obtained in many small fields, those fields must span several clustering scale lengths ($\sim 5h^{-1}$ Mpc comoving for typical galaxy samples, corresponding to a minimum field size of ~20 arcminutes in diameter at $z\sim 1$). The spectroscopic sample need not be representative in type or magnitude, but it must span the entire redshift range of and overlap spatially with the photometric sample that is to be calibrated.

Given these requirements, we now respond to specific questions from the committee:

**Q1. What O/IR capabilities are you using, are you planning to use, and will you need through the LSST era?** Sufficient spectroscopy for the most basic calibrations should be provided by the Baryon Acoustic Oscillation Experiment eBOSS (which began in Summer 2014). eBOSS should obtain redshifts of 1.5 million galaxies and QSO's over a redshift range of 0.6<z<4 over 7500 square degrees of sky. The planned Dark Energy Spectroscopic Instrument (DESI) will obtain redshifts of 20 million galaxies over a similar redshift range, but covering more than 14,000 square degrees of sky. DESI will have more than 3000 square degrees of overlap with LSST, vs. ~500 for eBOSS/LSST. DESI will therefore provide excellent calibration and redundant cross-checks via cross-correlations with different populations of objects at the same redshifts (e.g., both LRGs and ELGs from DESI can be used at z<1; both ELGs and QSOs at 1 < z < 1.6; and both QSOs and absorption-line systems or the Lyα forest at $z>1.6$). Redshifts from *Euclid* grism and/or *WFIRST/AFTA* grism and IFU spectroscopy could also contribute, but only in the limited redshift ranges for which the multiple line features needed for a secure redshift may be identified; hence ground-based O/IR spectroscopy is critical for this work.

The main DESI survey provides minimal numbers of redshifts at $z<0.6$; as seen in Fig. 3-3 of our Snowmass white paper, even with the incorporation of SDSS and BOSS redshifts, cross-correlation calibration falls short of LSST requirements at $z < 0.2$, which would mean that regime would need to be excluded from dark energy analyses, resulting in a modest degradation of cosmological constraints.  It was recently proposed that DESI be used during bright time that is not useful for the main survey to perform a survey of ~10 million galaxies with $r<19.5$ over 14k or 20k square degrees, which would provide a large spectroscopic sample with median redshift $z$~0.2.  ***If funded, this extension of DESI would enable accurate cross-correlation calibration over the full LSST redshift range***, in addition to enabling a variety of other science.

**Q2.  Do you have access to the O/IR capabilities you currently need to conduct your research?** There is some overlap between the membership of the eBOSS, DESI, and LSST teams; given that this overlap is not total, we expect that LSST will rely on the planned public archives of this data for our work.

**Q3.  Comment on the need for the U.S. community's access to non-federal O/IR facilities up to 30 meters in size:** If DESI does not go forward, it is likely that a large BAO survey would be pursued by scientists in other countries using one of a variety of proposed facilities (e.g. 4MOST or WEAVE).  In that scenario, LSST would require access to data from such surveys to obtain secure photo-z calibrations.

**Q10.  What types of scientific and observing coordination among the various NSF telescopes (including Gemini and LSST) and non-federal facilities are the most important for making scientific progress in the next 10-15 years?  How can such coordination best be facilitated?** Given LSST's location in the Southern hemisphere, an extension of DESI to the south via transfer to or duplication at the Blanco telescope would provide the best possible calibration from cross-correlations.  It would also enable many other synergies between imaging and spectroscopy, including science from the LSST imaging of spectroscopic objects directly; galaxy-galaxy lensing measurements; and constraints on the mapping between faint galaxies and dark matter halos provided by the same cross-correlation measurements used to calibrate photometric redshifts.  Such a transfer would also allow the measurement of spectroscopic redshifts for the majority of SN Ia hosts with well-sampled light curves from LSST, a project that is infeasible without the wide-field spectroscopic capabilities and high multiplexing afforded by DESI.